\documentclass[a4paper]{article}

\usepackage{INTERSPEECH2020}
\usepackage{ragged2e}
\usepackage{tipa}
\usepackage{tikz}
\usepackage{environ}
\usepackage{textcomp}
\usepackage{microtype}
\usepackage{booktabs}
\usetikzlibrary{calc} 
\usepackage{lipsum,adjustbox}
\usepackage{standalone}
\usepackage{subcaption}
\usepackage{tabularx}
\usepackage{balance}
\usepackage{cite}
\usepackage{amsmath}
\DeclareMathOperator{\clip}{clip}
\newcommand{\LeRF}{\textrm{LeRF}}
\newcommand{\MoRF}{\textrm{MoRF}}

\title{Large scale evaluation of importance maps in automatic speech recognition}
\name{Viet Anh Trinh${^1}$ and Michael I Mandel${^{1,2}}$}
\address{
  ${^1}$ The Graduate Center, CUNY, New York, USA\\
  ${^2}$ Brooklyn College, CUNY, New York, USA}
\email{vtrinh@gradcenter.cuny.edu, mim@sci.brooklyn.cuny.edu}

\begin{document}
\maketitle
\begin{abstract}
In this paper, we propose a metric that we call the structured saliency benchmark (SSBM) to evaluate importance maps computed for automatic speech recognizers on individual utterances. These maps indicate time-frequency points of the utterance that are most important for correct recognition of a target word. Our evaluation technique is not only suitable for standard classification tasks, but is also appropriate for structured prediction tasks like sequence-to-sequence models. Additionally, we use this approach to perform a large scale comparison of the importance maps created by our previously introduced technique using ``bubble noise'' to identify important points through correlation with a baseline approach based on smoothed speech energy and forced alignment. Our results show that the bubble analysis approach is better at identifying important speech regions than this baseline on 100 sentences from the AMI corpus.
\end{abstract}

\section{Introduction}

Finding relevant information in input features $X$ that is necessary for an output/task $y$ has seen a surge of interest in the computer vision \cite{zeiler2014visualizing,baehrens2010explain,smilkov2017smoothgrad,ribeiro2016should,ancona2017unified} and reinforcement learning communities \cite{mohamed2015variational,Goyal2020The,goyal2019infobot}. \cite{tishby2000information} proposed the information bottleneck approach to address the problem and \cite{achille2016information, alemi2016deep} used the idea to improve model generalization.  Our previous work proposed a correlational method to find regions of speech spectrograms that are important to a listener's correctly identifying the words it contains, and we applied it to both humans and automatic speech recognition (ASR) systems \cite{mandel16c,trinh20,trinh2018bubble}. These ``importance maps'' or ``saliency maps'' reveal how the ASR uses speech features to derive the recognition. In this paper, we propose a method to evaluate the quality of predicted importance maps and apply them to saliency maps estimated for an ASR ``listener.''


The saliency map in speech has a similar meaning to the saliency map in computer vision. However, unlike in vision, where ground truth can be obtained from eye-tracking systems, in speech, we do not have a corresponding `ear-tracking' system. With the human saliency map as the gold label, the predicted visual saliency map can be evaluated using different metrics, such as area under the curve (AUC) \cite{tatler2005visual}, correlation coefficient \cite{jost2005assessing},  similarity \cite{bylinskii2015saliency}, or information gain \cite{kummerer2015information}. For speech, where the human saliency map is unavailable, we propose a method to assess the quality of a predicted saliency map in this paper. The main idea of our approach is that the better the predicted saliency map, the higher the accuracy when the ASR uses only information from the important regions of the spectrogram. Similarly, if the important regions are removed from an observation, the ASR should have low accuracy. 

To the best of our knowledge, we are among the first, if not the first, to propose a method to evaluate the saliency map of running sentences, a structured prediction problem. In computer vision, there is related work on evaluation methods for saliency maps in simple classification problems without ground truth. \cite{samek2016evaluating} proposed the MoRF method (Most Relevant First) to evaluate saliency maps by measuring model performance degradation when the $n$ most relevant pixels are replaced by random values. \cite{ancona2017unified} introduced the complementary LeRF method (Least Relevant First), where the least relevant features are removed. \cite{Schulz2020Restricting} recommended evaluating with a score measuring the area between the MoRF and LeRF curves created when the number of pixels $n$ is varied. 

Inspired by \cite{samek2016evaluating,ancona2017unified,Schulz2020Restricting}, we propose here an evaluation metric, the SSBM, 
that measures accuracy degradation when the most or least important time-frequency points are replaced with white noise in a structured prediction setting. A fundamental difference between our approach and these others is that they evaluate the accuracy of a single simple classifier, such as an image classifier. Our approach, on the other hand, compares the recognition performance of a target word to that of the other words in the sentence.
We also characterize these values in comparison to the amount of speech energy preserved in a mixture.  

\section{Method}
The main idea of our method is to evaluate the quality of the predicted time-frequency importance regions for an utterance. Denote the predicted importance maps in the speech spectrogram from method $M$ for word $w$ as $I_M^w \in \{0,1\}^{F \times T}$, a binary matrix indicating whether time-frequency point $I_M^w(f,t)$ is important for the recognition of $w$ (1) or not (0). If the ASR can correctly recognize the speech using only the regions where $I_M^w = 1$ instead of using all the spectrogram points, and if it cannot recognize the speech when presented with only the regions where $I_M^w=0$, then we can conclude that method $M$ has successfully identified the important regions. To do this, we perform two tests. In the first case, we add noise everywhere in a sentence except the predicted important regions of $w$, which is equivalent to dropping the least relevant features (LeRF). In the second case, we add noise to the predicted important regions for $w$, equivalent to dropping the most relevant features (MoRF). If the ASR can recognize $w$ in the first case and not the second, then the importance region is correct. 
In addition, we also make sure that in these same two cases, the other words in the sentence are not correctly recognized when including the $I_M^{w}=1$ regions, and that the other words in the sentence are correctly recognized when excluding the $I_M^{w}=1$ regions.

We define a new metric that we call the structured saliency benchmark (SSBM) to evaluate the accuracy of the analyzed words with respect to the accuracy of other words in the sentence and the predicted important speech energies.
\begin{align}
    \Delta_{\LeRF} = \frac{a_{w} - a_{o}}{1-e_{\LeRF}} \qquad
    \Delta_{\MoRF} = \frac{a_o - a_w}{e_{\MoRF}}
    \end{align}
    \begin{align}
    SSBM = \Delta_{\LeRF} + \Delta_{\MoRF}
\end{align}
where $a_w$ is the accuracy of analyzed word $w$, $a_o$ is the averaged accuracy of the other words, $e_{\LeRF}$ is the percentage of the least important energy that is dropped (dropped energy is divided by utterance energy), and $e_{\MoRF}$ is the percentage of the most important energy that is dropped. Thus,  $\Delta_{\LeRF}$ represents the accuracy of the analyzed word per unit (percentage) of energy, with the accuracy of other words as a penalty. We can see that if the importance maps of $w$ are correct, then when the least important energy for $w$ is removed, the accuracy of $w$, $a_w$, should be high while the accuracy of other words, $a_o$, should be low. Additionally, for two different importance maps with the same $a_w$ and $a_o$, the map corresponding to higher $e_{\LeRF}$ (more unimportant energy is dropped) should be better as should the one with the lower  $e_{\MoRF}$ (less important energy is preserved).

In this paper, we analyze the importance maps of two different approaches. The first is a bubble analysis method where a time-frequency point is predicted to be important when its audibility in noise is significantly correlated with speech intelligibility \cite{mandel16c,trinh20}. The second is an energy-based baseline, where a time-frequency point in the spectrogram is predicted to be important when its energy is larger than a certain threshold.

\subsection{Bubble analysis approach}
The bubble analysis method \cite{mandel16c,trinh20} identifies important regions by adding many instances of random noise to clean speech, then finding the spectrogram points that are revealed when the ASR recognizes the noisy speech correctly and hidden by noise when the ASR fails to recognize the utterance. Specifically, the noisy utterances are generated by adding many instances of random white noise to the clean speech to make these utterances inaudible. However, the noise level is decreased inside randomly placed oval-shaped bubbles to reveal the speech information inside. Denote as $y_{ijk}$ the intelligibility, which has value one or zero (binary) when the ASR correctly or incorrectly recognizes the $k$th word in the $j$th noisy mixture of the $i$th clean utterance. In addition, the audibility $D_{ij}(f,t)$ is defined as a continuous variable that represents the inverse of the amount of noise added to a time-frequency point in a spectrogram, varying between zero (maximum noise) and one (no noise). A point-biserial correlation $c_{ik}(f,t)$ is computed between $D_{ij}(f,t)$ and $y_{ijk}$. The significance ($p$-value) of this correlation is examined under a two-sided t-test for every time-frequency point in the spectrogram \cite{trinh20}. The importance map is defined as the set of time-frequency points that have positive correlation and $p$-values smaller than a specific threshold.

\subsection{Energy-based approach}
In the energy-based approach, a time-frequency point in the spectrogram is considered important when its energy in a smoothed version of the spectrogram is greater than a certain threshold. Specifically, the linear frequency spectrogram has pre-emphasis applied, is converted to a mel spectrogram with 30 bins, and then is converted back to a linear frequency axis. The importance map of a word is then the set of high energy spectrogram points that are between the start and end frame of the target word in the forced alignment of the clean utterance produced by Kaldi.

\begin{figure}
\centering  
\includegraphics[width=1.0\columnwidth]{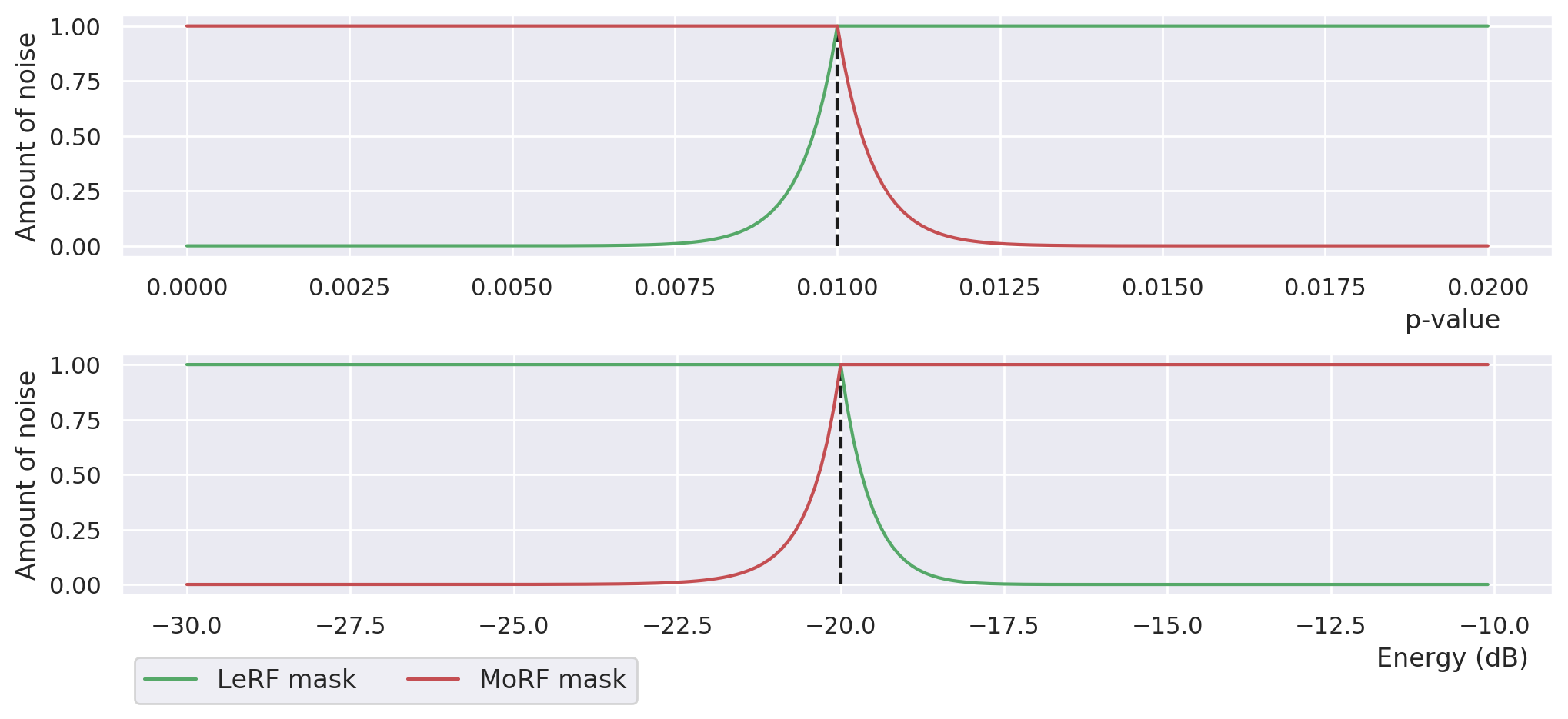}  
\caption{Example mask transition functions for an arbitrary threshold. Top: Bubble analysis. Bottom: Energy-based} 
\label{fig:maskThresholdSingle}
\end{figure}

\subsection{LeRF and MoRF noise masks}

The LeRF mask is created by adding maximum noise to unimportant regions while adding minimum noise to important regions. There is a transition between the two as shown in the top plot of Figure \ref{fig:maskThresholdSingle}. The intention is that when maximum noise is added outside the important regions of a specific word, then the ASR should still be able to recognize this word, but should not be able to recognize the other words in the sentence. The procedure is slightly different for the two mask prediction algorithms, so each is described separately below.

The bubble analysis LeRF mask $m_{\LeRF}^b$, at a single point is
\begin{align}
q_{\LeRF}^b(p) &= -(d_1 - d_0)\frac{p-t}{t\alpha - t}  \label{eq:lerfq1}
 \\
m_{\LeRF}^b(p) &= 10^{0.05 \clip(q_{\LeRF}^b(p),d_0,d_1)}
\label{eq:lerfq2}
\end{align}

where $t$ is the threshold, $p$ is the p-value of time-frequency points in the spectrogram, $\alpha < 1$ is a parameter controlling the size of the transition region while $d_0$ and $d_1$ control the minimum and maximum value of the mask, respectively.

\begin{figure} 
\centering  
\renewcommand{\arraystretch}{0.01}   \begin{tabular}{c} 
\includegraphics[width=\columnwidth]{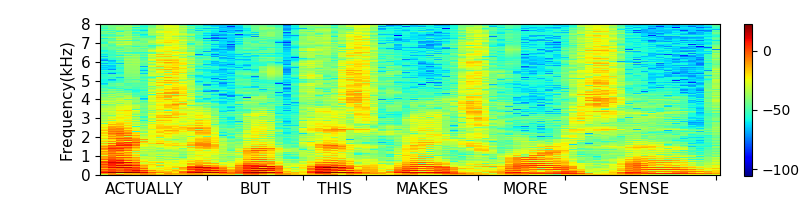} \\

\includegraphics[width=1.0\columnwidth]{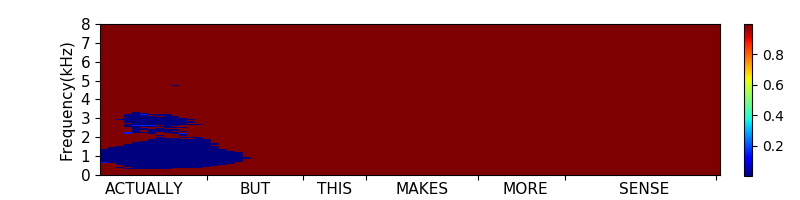}  \\
\includegraphics[width=1.0\columnwidth]{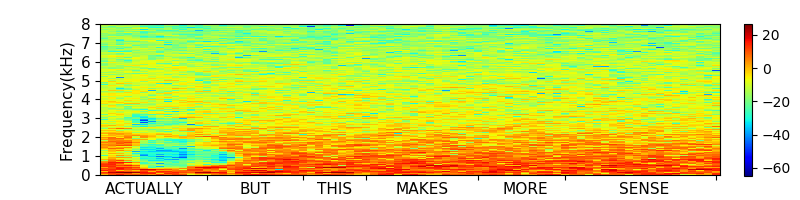}  \\
\end{tabular}
\caption{Bubble analysis approach. From top to bottom: 
(a) Clean speech
(b) LeRF mask created by dropping the least relevant features for the word ``actually'' with threshold $4.64 \times 10^{-7}$ (time-frequency points that have p-value $\geq 4.64 \times 10^{-7}$ have a maximum amount of noise added to them).
(c) Noisy mixture created by adding the mask in (b) to the clean speech in (a). 
} 
\label{fig:bubbleExceptMask}
\end{figure}

The green line in the top plot in Figure~\ref{fig:maskThresholdSingle} illustrates mask values for $t = 0.01$, $\alpha=0.5$. In addition, $d_0 =-80$, $d_1=0$ leading to a minimum mask value of $0.0001$ and maximum value of $1$.  As shown in this figure, a time-frequency point with a $p$-value larger than $0.01$ has noise level $1$ (maximum noise), while a point with a $p$-value smaller than $0.0075$ has noise level $0.0001$. Additionally, a visualization of a complete mask with threshold $t=4.64 \times 10^{-7}$ is shown in the second row of Figure~\ref{fig:bubbleExceptMask}.

\begin{figure}
\centering  
\renewcommand{\arraystretch}{0.05}     
\begin{tabular}{c} 
\includegraphics[width=1.0\columnwidth]{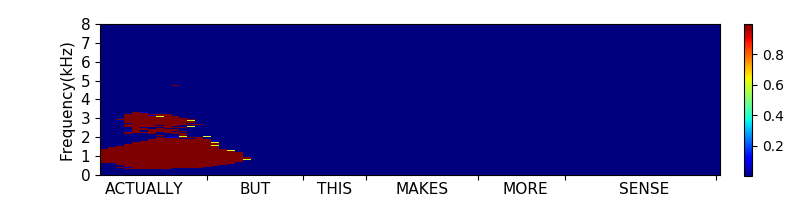}  \\
\includegraphics[width=1.0\columnwidth]{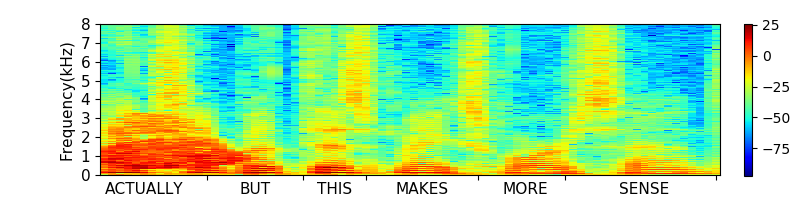}  \\
\end{tabular}
\caption{Bubble analysis. Top: MoRF mask created by dropping the most important features of the word ``actually'' with threshold $ 4.64 \times 10^{-7}$. Bottom: Noisy mixture} 
\label{fig:onlyThreshold}
\end{figure}

The bubble analysis MoRF mask is derived in a similar way as equations \eqref{eq:lerfq1} and \eqref{eq:lerfq2}, however with  $q_{\MoRF}^b(p) = -q_{\LeRF}^b(p)$.
The red line in the top plot of Figure \ref{fig:maskThresholdSingle} shows the MoRF mask with the same parameters as the green line.  Besides, visualization of the mask are shown in the top plot of Figure \ref{fig:onlyThreshold}. In general, we can observe that the MoRF and LeRF mask are almost complementary to each other, but are not exactly because we always have the masks provide decay smoothly towards 0 to mirror the logarithmic nature of intensity perception.

\begin{figure}
\centering  
\renewcommand{\arraystretch}{0.05}
\begin{tabular}{c} 
\includegraphics[width=1.0\columnwidth]{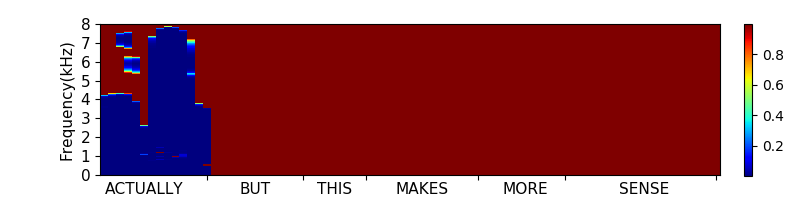}  \\
\includegraphics[width=1.0\columnwidth]{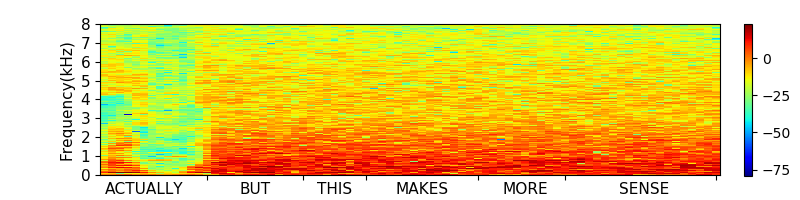}  \\
\end{tabular}
\caption{Energy-based approach. 
Top: LeRF mask with $ t_{dB}=-20$. 
Bottom: Noisy mixture
} 
\label{fig:baseExceptMask}
\end{figure}

Similarly, the LeRF mask for the energy-based approach is created by adding maximum noise to the time-frequency region with energy lower than a specific threshold $t_{\textrm{dB}}$ in decibels (unimportant regions). The important regions have minimum noise added, except the transition area. The mask $m$ is defined as
\begin{align}
q_{\LeRF}^e(a) &= (d_1 - d_0)\frac{a- t}{\alpha t - t}  \label{eq:lerfbase2}  \\
m_{\LeRF}^e(a) &= 10^{0.05 \clip(q_{\LeRF}^e(a),d_0,d_1)}
\label{eq:lerfbase3}
\end{align}
where $a$ is the absolute magnitude of the time-frequency point in the spectrogram and $t =10^{0.05t_{\textrm{dB}}}$ is the threshold in magnitude. An example of the mask with a specific threshold $t_{\textrm{dB}}=-20$~dB is illustrated in the bottom plot of Figure~\ref{fig:maskThresholdSingle} and Figure~\ref{fig:baseExceptMask}.

The energy-based MoRF mask is formed by adding maximum noise to the time-frequency region with energy bigger than a specific threshold. The mask is derived the same as equations \eqref{eq:lerfbase2} and \eqref{eq:lerfbase3} except with $q_{\MoRF}^e(a) = -q_{\LeRF}^e(a)$.


To create the noisy speech, we multiply the spectrogram of a white noise signal by the mask and add the masked noise to the clean speech.  Examples of the mask and the masked noisy speech are shown in the second and third rows of Figure~\ref{fig:bubbleExceptMask}.

\section{Experimental setup}

We utilize the AMI dataset~\cite{carletta2005ami}, which includes 100 hours of English meeting recordings. We selected the Individual Headset Microphone (IHM) channels for our experiment. We followed the standard train/test split and chose 100 sentences from the test set where the recognizer achieved 100\% accuracy without additional noise added to be our set of clean speech. We created 1000 noisy mixtures for every clean utterance, leading to a dataset of 100,000 mixtures for the bubble analysis method.

\begin{figure*}
     \includegraphics[width=\textwidth]{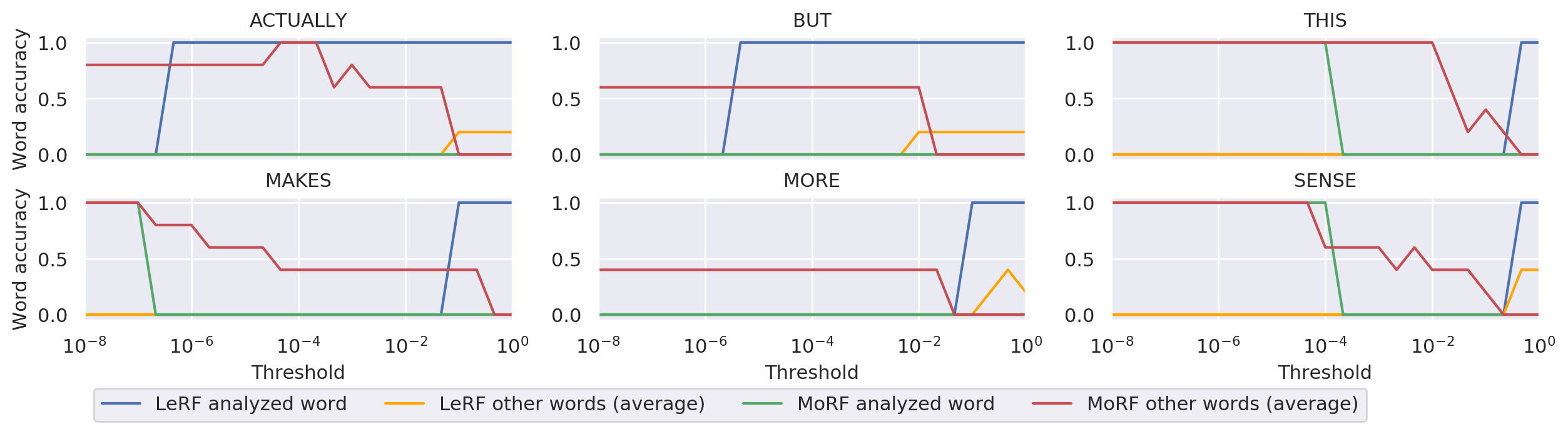}
     \caption{Bubble analysis: Word accuracy on the sentence ``actually but this makes more sense.'' with LeRF and MoRF masks} \label{fig:bubblesentence85}
\end{figure*}
We use Kaldi~\cite{PoveyEtAl2011} as the ASR to perform the experiments. We choose the standard model in AMI recipe s5b with a time-delay neural network (TDNN) acoustic model and an $n$-gram language model from the SRI Language Modeling Toolkit (SRILM)~\cite{stolcke2002srilm}. The TDNN is a modification of a feed-forward neural network, where the hidden vector representation at a layer is derived from several vectors (window of size $n$) from the preceding layer. The time-domain utterances are sampled at 16~kHz and are transformed into the frequency domain using an STFT with window length 64~ms, and hop length 16~ms.

For the bubble analysis technique, we choose $d_0=-80$, $d_1=0$ and $\alpha = 0.5$. We perform experiments with 25 different values of threshold $t$ that are spaced evenly on a log scale from $10^{-8}$ to $10^0$. 
For the energy-based technique, we use the same values of $d_0,d_1,\alpha$, however we use 21 different values of thresholds $t_{dB}$, spaced evenly on a linear scale from $-80$ to $20$ with a step size of $5$.

\section{Results}

\begin{figure}
\centering  
\includegraphics[width=1.0\columnwidth]{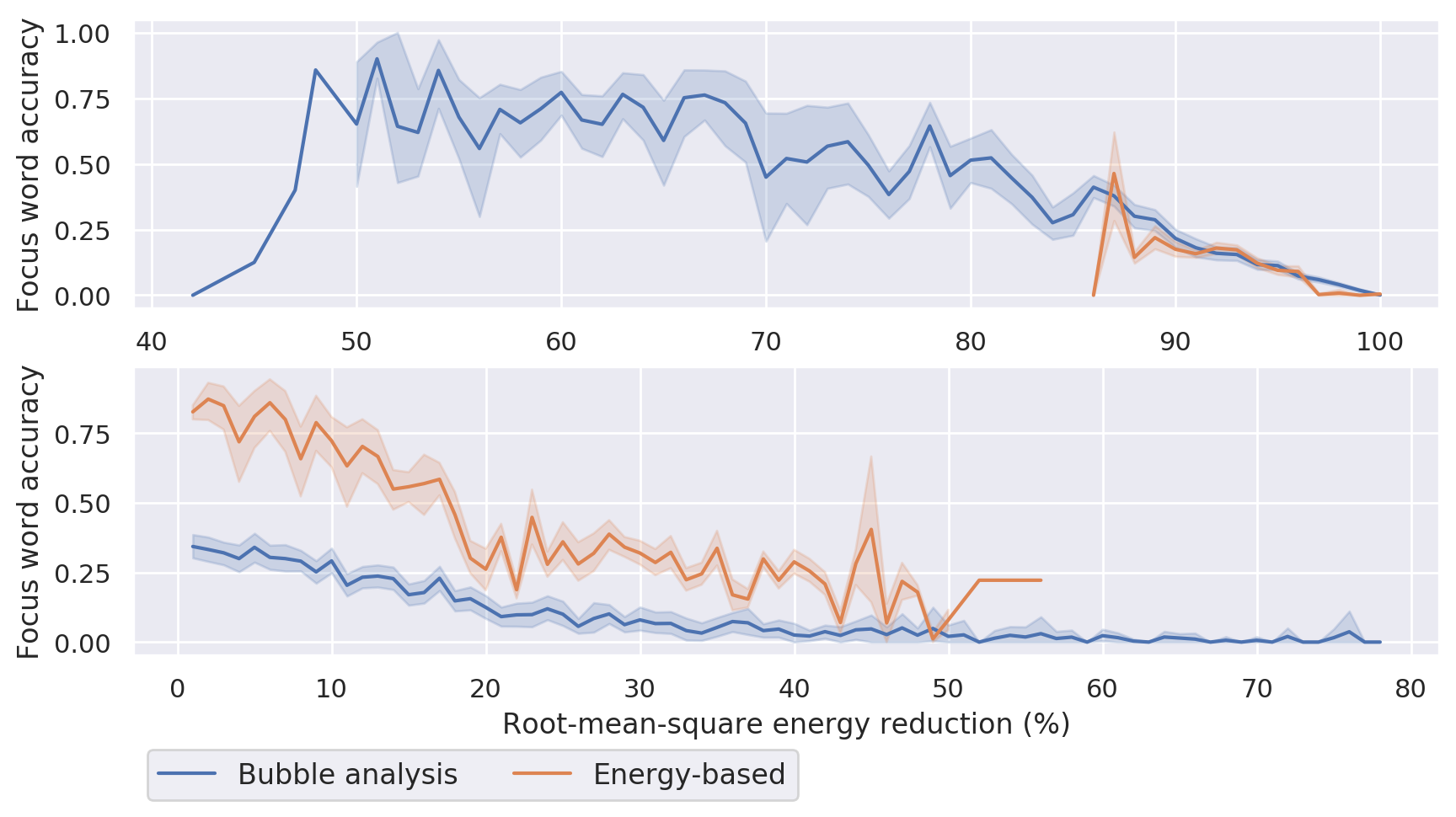}
\caption{Average accuracy of analyzed word with LeRF mask (top) and MoRF (bottom). 
} 
\label{fig:percent}
\end{figure}

Here, we compare the bubble analysis and the energy-based approaches according to LeRF and MoRF curves and SSBM scores. 
Figure~\ref{fig:percent} allows a direct comparison between the two mask methods by characterizing each masked signal by the proportion of speech energy in the entire utterance that it obscures. This proportion could vary for different words at the same threshold, so this plot averages over masks that have the same proportion when rounded to the nearest percent. 

The top plot of Figure~\ref{fig:percent} shows the accuracy of analyzed words when the least important features are dropped, averaged over the entire dataset. Perfect performance in this case would be in the top right corner, obscuring almost all of the speech while preserving recognition accuracy.
In general, we can see that the bubble analysis method (blue line) achieves slightly higher accuracy than the energy-based method (orange line). 
For instance, by dropping the $89\%$ least important energy, the energy-based and bubble analysis masks have average word accuracy $22\%$  and $29\%$, respectively.

The bottom plot of Figure \ref{fig:percent}
shows the accuracy of analyzed words when the most important features are dropped on all 100 sentences. A perfect MoRF mask would be in the bottom left corner of the bottom plot, obscuring almost none of the speech while destroying recognition accuracy. This plot demonstrates that the bubble analysis method is better at reducing recognition accuracy than the energy-based method when both drop the same amount of important speech energy. In both plots, the orange lines are shorter than the blue lines because the important regions of a word are restricted to be between the start frame and end frame in the energy-based approach.


\begin{figure}[t]
\centering  
\includegraphics[width=1.0\columnwidth]{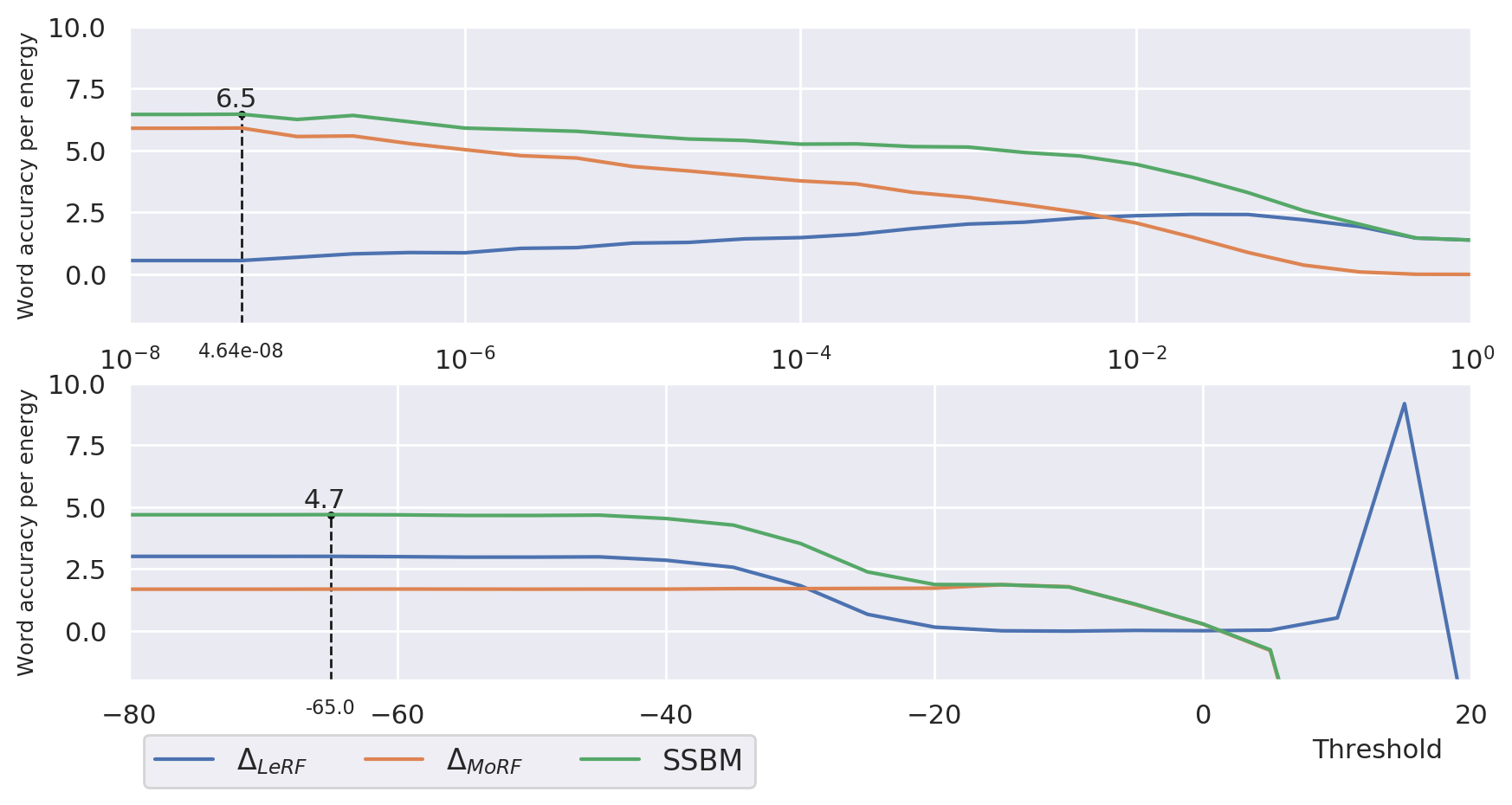}  
\caption{$\Delta_{\LeRF}$ and $\Delta_{\MoRF}$ along with their combination into the SSBM score. Higher is better for all three. Top: bubble analysis, achieving SSBM of 6.5. Bottom: energy-based, achieving SSBM of 4.7 (accuracy per unit (percentage) of
energy).}
\label{fig:SSBM}
\end{figure}

\begin{figure}
     \includegraphics[width=\columnwidth]{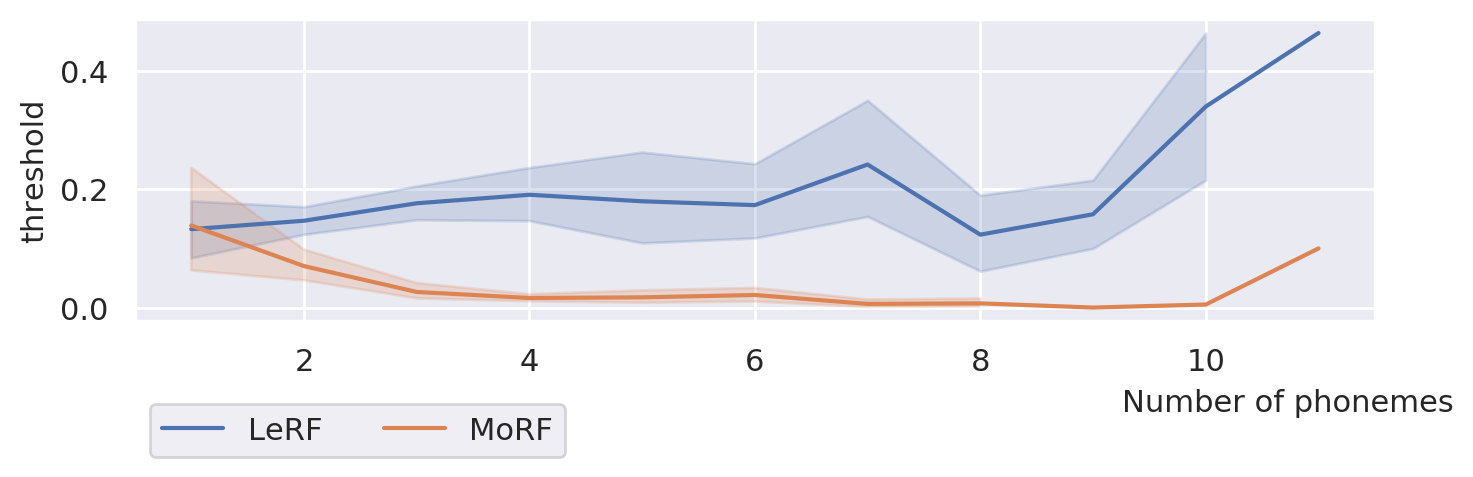}
     \caption{Relationship between number of phonemes and threshold for the bubble analysis mask.} \label{fig:phoneme}
\end{figure}

Figure \ref{fig:SSBM} shows the SSMB scores (green lines) at various thresholds for both methods. For the bubble analysis method in the top row, we can see that the threshold of $4.64 * 10^{-8}$ obtains the best SSBM score of 6.5. This means that the increase in LeRF accuracy at higher thresholds is not worth the decrease in MoRF accuracy. For the energy-based method in the second row, the threshold of $-65$~dB achieves the highest SSMB score of 4.7, which is worse than that of the bubble analysis method. Thus, the bubble analysis method produces better importance maps than the energy-based approach according to the LeRF and MoRF curves and the SSBM score.

\subsection{Other discussions}

First, we can see that the ASR does not need to observe all of the speech energy of a word to correctly identify it. For illustration, the ASR can recognize the word ``actually'' with a threshold as low as $4.64 \times 10^{-7}$ on bubble analysis LeRF mask as in Figure~\ref{fig:bubblesentence85} (blue line). This mask and its corresponding noisy speech are illustrated in the second and third row of Figure~\ref{fig:bubbleExceptMask}. As we can see, the mask only spans from 400~Hz to 3200~Hz. Surprisingly, the clean speech lacks energy at those frequencies, but this does not prevent the ASR from correctly identifying the word. 
 
Second, the threshold identifying which time-frequency points are important is varied across word. 
For example, in Figure~\ref{fig:bubblesentence85} (blue line), the ASR needs to use all time-frequency points with $p$-value $ < 4.64 \times 10^{-6}$ to correctly identify the word ``but'', however, the ASR must use all spectrogram points with $p$-value $< 0.1$  to recognize the words ``more''. 

Figure~\ref{fig:phoneme} shows a possible explanation of why the threshold for correct recognition varies across words. It shows the threshold at which a target words transitions from correct to incorrect recognition as a function of word length in phonemes. We can see that longer words typically require a higher LeRF threshold, meaning more speech is revealed, while they typically require a lower MoRF threshold, meaning less speech is obscured. Similar trends were observed with word length measured in syllables and characters.

\section{Conclusion and future work}
In this paper, we proposed an evaluation metric for structured saliency maps, where we measure the word accuracy when either keeping or dropping the most important regions. A gap in this accuracy is measured between the analyzed word and other words in the sentence with respect to the predicted important speech energies. Additionally, we perform a large scale saliency map analysis with a bubble analysis method and energy-based baseline on sentences from the AMI meeting corpus. According to several metrics, the bubble analysis approach achieves a better importance map than its alternative. In the future, we will extend this evaluation to different methods to compare different speech importance maps and use these importance maps to enhance speech recognition robustness in noisy conditions. We also hope that this speech saliency evaluation metric can facilitate a community evaluation on the topic of speech saliency, similar to those that have been organized around visual saliency~\cite{mit-tuebingen-saliency-benchmark}. 

\balance
\bibliographystyle{IEEEtran}
\bibliography{mybib}
\end{document}